\documentclass[aps,prl,reprint,showpacs,superscriptaddress,citeautoscript]{revtex4-1}
\usepackage{lmodern}
\usepackage{amsmath}
\usepackage{amssymb}
\usepackage{textcomp}
\usepackage{gensymb}
\usepackage{graphicx}
\usepackage{wasysym}
\usepackage{units}
\usepackage{array}
\usepackage{multirow}
\usepackage[bookmarks=false]{hyperref}
\hypersetup{
		breaklinks=true,
    unicode=false,
    pdftoolbar=true,
    pdfmenubar=true,
    pdffitwindow=false,
    pdfstartview={FitH},
    pdftitle={Taufour},
    pdfauthor={Taufour},
    pdfsubject={LaCrGe3},
    pdfcreator={Taufour},
    pdfproducer={Taufour},
    pdfkeywords={keyword1} {key2} {key3},
    pdfnewwindow=true,
    colorlinks=true,
    linkcolor=black,
    citecolor=black,
    filecolor=black,
    urlcolor=blue
}

\begin{document}

\title{Ferromagnetic Quantum Critical Point Avoided by the Appearance of Another Magnetic Phase in LaCrGe$_3$ under Pressure}
\author{Valentin \surname{Taufour}}
\email{vtaufour@ucdavis.edu}
\affiliation{The Ames Laboratory, US Department of Energy, Iowa State University, Ames, Iowa 50011, USA}
\author{Udhara S. \surname{Kaluarachchi}}
\affiliation{The Ames Laboratory, US Department of Energy, Iowa State University, Ames, Iowa 50011, USA}
\affiliation{Department of Physics and Astronomy, Iowa State University, Ames, Iowa 50011, U.S.A.}%
\author{Rustem \surname{Khasanov}}
\affiliation{Laboratory for Muon Spin Spectroscopy, Paul Scherrer Institute, CH-5232 Villigen PSI, Switzerland}%
\author{Manh Cuong \surname{Nguyen}}
\affiliation{The Ames Laboratory, US Department of Energy, Iowa State University, Ames, Iowa 50011, USA}
\author{Zurab \surname{Guguchia}}
\affiliation{Laboratory for Muon Spin Spectroscopy, Paul Scherrer Institute, CH-5232 Villigen PSI, Switzerland}%
\author{Pabitra Kumar \surname{Biswas}}
\affiliation{ISIS Pulsed Neutron and Muon Source, STFC Rutherford Appleton Laboratory, Harwell Campus, Didcot, Oxfordshire, OX11 0QX, United Kingdom}%
\author{Pietro \surname{Bonfà}}
\affiliation{Dipartimento di Fisica e Scienze della Terra, Parco Area delle Scienze 7/A, I-43124 Parma, Italy}%
\author{Roberto \surname{De Renzi}}
\affiliation{Dipartimento di Fisica e Scienze della Terra, Parco Area delle Scienze 7/A, I-43124 Parma, Italy}%
\author{Xiao \surname{Lin}}
\affiliation{Department of Physics and Astronomy, Iowa State University, Ames, Iowa 50011, U.S.A.}%
\author{Stella~K. \surname{Kim}}
\affiliation{The Ames Laboratory, US Department of Energy, Iowa State University, Ames, Iowa 50011, USA}
\affiliation{Department of Physics and Astronomy, Iowa State University, Ames, Iowa 50011, U.S.A.}%
\author{Eun~Deok \surname{Mun}}
\affiliation{Department of Physics and Astronomy, Iowa State University, Ames, Iowa 50011, U.S.A.}%
\author{Hyunsoo \surname{Kim}}
\affiliation{Department of Physics and Astronomy, Iowa State University, Ames, Iowa 50011, U.S.A.}%
\author{Yuji \surname{Furukawa}}
\affiliation{The Ames Laboratory, US Department of Energy, Iowa State University, Ames, Iowa 50011, USA}
\affiliation{Department of Physics and Astronomy, Iowa State University, Ames, Iowa 50011, U.S.A.}%
\author{Cai-Zhuang \surname{Wang}}
\affiliation{The Ames Laboratory, US Department of Energy, Iowa State University, Ames, Iowa 50011, USA}
\author{Kai-Ming \surname{Ho}}
\affiliation{The Ames Laboratory, US Department of Energy, Iowa State University, Ames, Iowa 50011, USA}
\affiliation{Department of Physics and Astronomy, Iowa State University, Ames, Iowa 50011, U.S.A.}%
\author{Sergey~L. \surname{Bud'ko}}
\affiliation{The Ames Laboratory, US Department of Energy, Iowa State University, Ames, Iowa 50011, USA}
\affiliation{Department of Physics and Astronomy, Iowa State University, Ames, Iowa 50011, U.S.A.}%
\author{Paul~C. \surname{Canfield}}
\affiliation{The Ames Laboratory, US Department of Energy, Iowa State University, Ames, Iowa 50011, USA}
\affiliation{Department of Physics and Astronomy, Iowa State University, Ames, Iowa 50011, U.S.A.}%

\begin{abstract}
The temperature-pressure phase diagram of the ferromagnet LaCrGe$_3$ is determined for the first time from a combination of magnetization, muon-spin-rotation and electrical resistivity measurements. The ferromagnetic phase is suppressed near $2.1$~GPa, but quantum criticality is avoided by the appearance of a magnetic phase, likely modulated, AFM$_Q$. Our density functional theory total energy calculations suggest a near degeneracy of antiferromagnetic states with small magnetic wave vectors $Q$ allowing for the potential of an ordering wave vector evolving from $Q=0$ to finite $Q$, as expected from the most recent theories on ferromagnetic quantum criticality. Our findings show that LaCrGe$_3$ is a very simple example to study this scenario of avoided ferromagnetic quantum criticality and will inspire further study on this material and other itinerant ferromagnets.
\end{abstract}

\maketitle

Systems with a quantum phase transition (QPT), a phase transition that occurs at $0$~K, have revealed a wide variety of enigmatic phenomena. The case of the paramagnetic to ferromagnetic (PM-FM) QPT itself can lead to various phase diagrams with the occurrence of tricritical wings~\cite{Uemura2007Nature,Levy2007NP,Taufour2010PRL}, a quantum Griffiths region~\cite{Westerkamp2009PRL,UbaidKassis2010PRL}, superconductivity~\cite{Saxena2000Nature,Aoki2001Nature,Huy2007PRL,Levy2007NP,Cheng2015PRL}, or non-Fermi liquid behavior~\cite{Pfleiderer2001Nature}. Several theories have been proposed to explain these intriguing effects~\cite{Vojta2003PRL,Belitz2005PRL,Hoyos2008PRL,Kruger2014PRL}. Current theoretical proposals suggest that a continuous PM-FM QPT is not possible in clean, fully ordered, systems. Instead, the transition can be of the first order, or a modulated magnetic phase can appear~\cite{Belitz1997PRB,Chubukov2004PRL,Conduit2009PRL,Karahasanovic2012PRB,Thomson2013PRB,Pedder2013PRB,AbdulJabbar2015NP}. In this Letter, we identify a new system, the compound LaCrGe$_3$, where a continuous PM-FM QPT under pressure is avoided by the appearance of a magnetic phase, most likely a modulated phase characterized by a small wave vector $Q$. LaCrGe$_3$ provides a clean example of a simple $3d$-shell transition metal system in which such a phase appears.

Long-wavelength correlation effects are essential to the appearance of the modulated magnetic phase (AFM$_Q$) which is, therefore, characterized by a small wave vector $Q$~\cite{Belitz1997PRB}. In order to study such phases experimentally, it is necessary to identify a system with a FM state that can be tuned to a QPT using a clean tuning parameter. When chemical substitutions are used to drive the PM-FM transition to $0$~K, defects (and sometimes even changes in band filling) are inevitably introduced. Such quenched disorder is expected to smear the QPTs~\cite{Hoyos2008PRL}. Pressure is considered as one of the cleanest parameters to tune a system towards a QPT but usually limits the number of experimental techniques that can be used to probe an eventual new phase. In this study, resistivity measurements are used for a precise mapping of the temperature-pressure phase diagram of LaCrGe$_3$. We combine these with magnetization and muon-spin-rotation ($\mu$SR) measurements that probe the new phase AFM$_{Q}$ and show that AFM$_{Q}$ has a similar magnetic moment as the FM phase but without net macroscopic magnetization. Finally, using thermodynamic considerations as well as total energy calculations, we show that there is a near degeneracy of AFM ordered states near $2$~GPa that can allow for the evolution of an ordering wave vector from $Q=0$ to $Q>0$. Taken together, these data firmly establish LaCrGe$_3$ as a clear example of avoided ferromagnetic quantum criticality by the appearance of modulated magnetic phase.

LaCrGe$_3$ crystallizes in the hexagonal BaNiO$_3$-type structure [space group \textit{P}6$_3$/\textit{mmc} (194)]~\cite{Bie2007CM,Cadogan2013SSP}. At ambient pressure, LaCrGe$_3$ is ferromagnetic below the Curie temperature $T_{\textrm{C}}=85$~K~\cite{Lin2013PRB} with an ordered magnetic moment at low temperature of $1.25$~$\mu_B$/Cr aligned along the $c$~axis~\cite{Cadogan2013SSP,Lin2013PRB}. This rather small value of the magnetic moment compared with the effective moment above $T_{\textrm{C}}$ ($\mu_{\textrm{eff}}=2.4$~$\mu_B$/Cr) suggests some degree of delocalization of the magnetism~\cite{Rhodes1963PRSLA,Lin2013PRB} in agreement with band-structure calculations~\cite{Bie2007CM}.

\begin{figure}[!htb]
\begin{center}
\includegraphics[width=7.2cm]{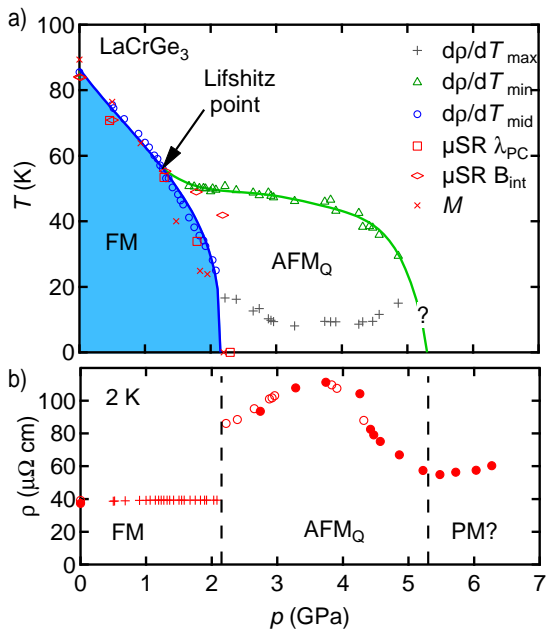}
\caption{\label{fig:diagab}(a)~Temperature-pressure phase diagram of LaCrGe$_3$ from various measurements showing the ferromagnetic (FM) and modulated magnetic phase (AFM$_Q$). (b)~Pressure dependence of the resistivity at $2$~K. Different symbols (cross, open circle, filled circle) represent data of different samples from different pressure runs.}
\end{center}
\end{figure}

Figure~\ref{fig:diagab}(a) shows the temperature-pressure phase diagram of LaCrGe$_3$ obtained from resistivity, magnetization, and $\mu$SR measurements which will be described below. The ferromagnetic phase is suppressed at $p=2.1$~GPa and a modulated magnetic phase labeled AFM$_Q$ is observed for $1.5<p<5.3$~GPa. The very steep pressure dependence of $T_{\textrm{C}}$ near $2.1$~GPa and the abrupt doubling of the residual ($T=2$~K) resistivity shown in Fig.~\ref{fig:diagab}(b) suggest that the FM-AFM$_{Q}$ transition is of the first order. Indeed, the Clausius-Clapeyron relation imposes that $dT_{\textrm{C}}/dp$ tends to infinity for a first-order transition at $T=0$~K, and a peak in the resistivity rather than a discontinuous step is expected for a second-order quantum phase transition~\cite{Miyake2002JPSJ,Raymond2000JLTP,Knebel2008JPSJ}. The discontinuous step disappears near $40$~K~\cite{*[{See the Supplemental Material attached, which includes Refs.~\cite{Budko1984ZETF,Voronovsky1979ZETF,Kim2011PRB,Colombier2007RSI,Tateiwa2009RSI,Piermarini1973JAP,Bireckoven1988JPESI,Alireza2007JPSJS,Murata2008RSI,Piermarini1975JAP,Khasanov2016HPR,Eiling1981,Springer2015-180,Maisuradze2011PRB,Kirkpatrick2015PRL,Brando2016RMP,Kohn1965PR,Perdew1981PRB,Blochl1994PRB,Kresse1999PRB,Kresse1996CMS,Monkhorst1976PRB}, with additional details on experimental methods and calculations and additional $\mu$SR and resistivity data}][{}] SupplMat}\nocite{Budko1984ZETF,Voronovsky1979ZETF,Kim2011PRB,Colombier2007RSI,Tateiwa2009RSI,Piermarini1973JAP,Bireckoven1988JPESI,Alireza2007JPSJS,Murata2008RSI,Piermarini1975JAP,Khasanov2016HPR,Eiling1981,Springer2015-180,Maisuradze2011PRB,Kirkpatrick2015PRL,Brando2016RMP,Kohn1965PR,Perdew1981PRB,Blochl1994PRB,Kresse1999PRB,Kresse1996CMS,Monkhorst1976PRB} above which the resistivity isotherms are a continuous function of pressure. Therefore, there exists a tricritical point near $40$~K below which the FM-AFM$_Q$ transition is of the first order. The merging of the three transition lines (FM-AFM$_Q$, FM-PM, and AFM$_Q$-PM) is called the Lifshitz point~\cite{Hornreich1975PRL}.

\begin{figure}[!htb]
\begin{center}
\includegraphics[width=5.1cm]{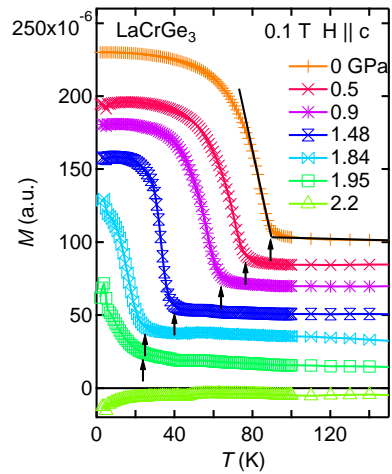}
\caption{\label{fig:Mag1000Oe}Temperature dependence of the magnetization at various pressures. The Curie temperature is indicated by arrows at the position of the change of slope as illustrated by lines for the curve at $0$~GPa. Data are offset for clarity. Because such measurement is sensitive to the large background of the pressure cell with comparison to the sample signal, the applied field is rather low ($0.1$~T). Since at such low field the magnetization does not reach saturation, the change of magnetization at low temperature is better determined by $\mu$SR experiments.}
\end{center}
\end{figure}

The suppression of the FM phase with applying pressure can be seen directly from the magnetization measurements (Fig.~\ref{fig:Mag1000Oe}). The FM phase is revealed by a sharp increase of the magnetization upon cooling below $T_{\textrm{C}}$. The pressure evolution of the transition can be followed up to $1.95$~GPa. At $2.2$~GPa, no FM transition can be observed.

\begin{figure}[!htb]
\begin{center}
\includegraphics[width=8.3cm]{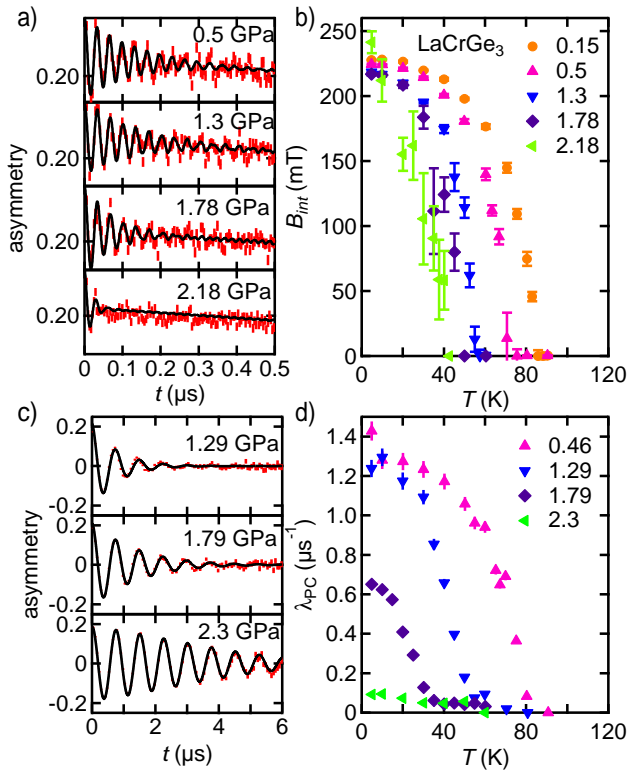}
\caption{\label{fig:muonsabcd}(a)~Zero field $\mu$SR spectra at $5$~K and various pressures. The solid lines are fits using Eqs.~(1)-(3) in Ref.~\cite{SupplMat}. (b)~Temperature dependence of the internal field at various pressures. (c)~$\mu$SR spectra at $5$~K in a weak transverse field $\mu_0H=10$~mT at various pressures. The solid lines are fits by using Eq.~(4) in Ref.~\cite{SupplMat}. (d)~Temperature dependence of the pressure cell relaxation $\lambda_{\textrm{PC}}$ at various pressures.}
\end{center}
\end{figure}

A similar decrease of $T_{\textrm{C}}$ is observed in the $\mu$SR experiments. The $\mu$SR spectra obtained in zero field at $5$~K are shown in Fig.~\ref{fig:muonsabcd}(a) from which we obtain the internal field at the muon site $B_\textrm{int}$ [Fig.~\ref{fig:muonsabcd}(b)]. Another set of spectra shown in Fig.~\ref{fig:muonsabcd}(c) is measured with a weak transverse field $\mu_0H=10$~mT from which the relaxation from the pressure cell contribution $\lambda_\textrm{PC}$ was obtained [Fig.~\ref{fig:muonsabcd}(d)]. For $p<1.4$~GPa, a simultaneous increase of $B_\textrm{int}$ and $\lambda_\textrm{PC}$ can be observed upon cooling through $T_\textrm{C}$, indicating that the sample is ferromagnetic and induces a field in the pressure cell body. At $1.78(1)$~GPa, the increase is no longer simultaneous, which corresponds to the pressure range where the AFM$_Q$ phase has a higher transition temperature than $T_\textrm{C}$. Above $2.1$~GPa, the ferromagnetic phase is completely suppressed: no field is induced around the sample so that no additional depolarization of the muon spin polarization from the cell body is measured [Figs.~\ref{fig:muonsabcd}(c) and \ref{fig:muonsabcd}(d)]. However, the AFM$_Q$ phase transition is still observed as an increase of $B_\textrm{int}$ upon cooling [Fig.~\ref{fig:muonsabcd}(b)], demonstrating that the AFM$_Q$ phase is magnetic in nature. It is important to note that the internal field at the muon site at low temperatures does not change significantly with pressure. In particular, it is nearly unchanged between the FM and the AFM$_Q$ phase. This is consistent with the AFM$_Q$ phase having similar ferromagnetic planes as the FM phase but with a modulation (wave vector $Q$) so that there is no macroscopic magnetization.

\begin{figure}[!htb]
\begin{center}
\includegraphics[width=8.4cm]{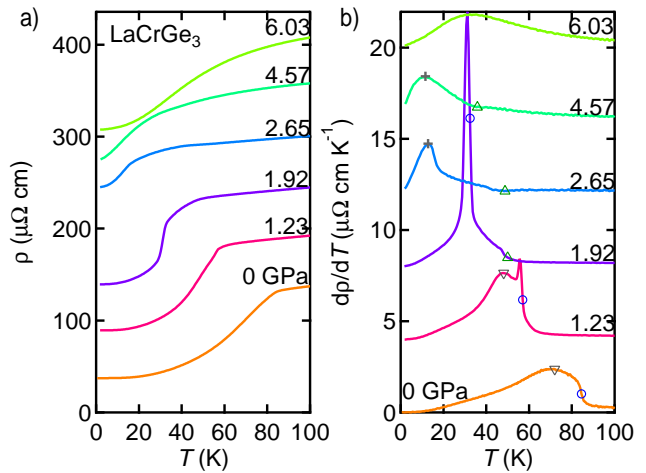}
\caption{\label{fig:rodrodTab}(a)~Temperature dependence of the electrical resistivity at various pressures. Data are offset by $50$~$\mu\Omega$~cm for clarity. (b)~Temperature dependence of the temperature derivative of the resistivity $d\rho/dT$ at various pressures. Data are offset by $4$~$\mu\Omega$~cm~K$^{-1}$ for clarity. Symbols represent the position of the anomalies.}
\end{center}
\end{figure}

We show the temperature dependence of the resistivity of LaCrGe$_3$ at representative pressures in Fig.~\ref{fig:rodrodTab}(a) and the corresponding temperature derivatives in Fig.~\ref{fig:rodrodTab}(b). The PM-FM transition, which later becomes the AFM$_Q$-FM transition, is revealed as a sharp increase in $d\rho/dT$ upon cooling (e.g., at $0$~GPa), which progressively evolves into a sharp peak (e.g., at $1.23$~GPa). Below the PM-FM transition at $0$~GPa, a broad maximum is observed in $d\rho/dT$ [gray triangle in Fig.~\ref{fig:rodrodTab}(b)], whereas no corresponding anomaly can be observed in magnetization (Fig.~\ref{fig:Mag1000Oe}), internal field (Fig.~\ref{fig:muonsabcd}), or specific heat~\cite{Lin2013PRB}. This may correspond to a crossover within the ferromagnetic state as observed in the superconducting ferromagnet UGe$_2$~\cite{Taufour2010PRL,Pfleiderer2002PRL}. The PM-AFM$_Q$ transition is revealed as a small bump in $\rho(T)$ upon cooling, which is better seen as a kink in $d\rho/dT$ (e.g., at $1.92$~GPa and at $2.27$~GPa in Fig.~SI.3~\cite{SupplMat}). This is often characteristic of a modulated magnetic order (spin-density wave) as a small gap in the density of states opens around a nesting vector of the Fermi surface. With the disappearance of the AFM$_Q$-FM transition above $2.1$~GPa, a sharp decrease in $\rho(T)$ upon cooling is observed (another peak in $d\rho/dT$) that probably indicates a transition to a state with another $Q$ vector in the AFM$_{Q}$ phase (e.g., at $2.65$~GPa). Above $3.7$~GPa, this sharp decrease seems to become broader (e.g., at $4.57$~GPa). These other anomalies within the AFM$_Q$ phase are compatible with a temperature and pressure dependence of the wave vector $Q$. The PM-AFM$_Q$ transition can still be seen at $4.57$~GPa but not at $6$~GPa [Fig.~\ref{fig:rodrodTab}(b)], and we estimate that the AFM$_Q$ phase is suppressed around $5.3$~GPa where a minimum is also observed in the low-temperature resistivity [Fig.~\ref{fig:diagab}(b)].

We now discuss the implication of the very peculiar phase diagram of LaCrGe$_3$. When the FM-AFM$_Q$ transition is of the second-order near the Lifshitz point, the order parameter must change continuously between these two phases; i.e., the magnetization $M_0$ vanishes, whereas the staggered magnetization $M_Q$ increases continuously. A possibility is to have the magnetic $Q$ vector varying continuously at the transition from $Q=0$ to $Q>0$. In essence, the internal field would not change on a few~\AA~ level, but, as soon as $Q$ becomes finite, the average of the magnetization over the magnetic unit cell becomes zero. In this case, the $Q$~vector would take very small values near the FM-AFM$_Q$ transition and grow to larger values deeper in the AFM$_Q$ phase. Within the AFM$_Q$ region of the phase diagram, the $Q$~vector can change with pressure and temperature, but at a given temperature and pressure value, the sample has a given $Q$~vector. As the system evolves away from the Lifshitz point, it is likely that the $Q$~vector will ``lock" at some specific values at low temperature which is consistent with a transition to other phases with different wave vectors in the AFM$_Q$ region. A finite $Q$~vector is also consistent with the first-order nature of the FM-AFM$_{Q}$ transition at low temperature. Although it can be very difficult to measure small-$Q$ wave vectors at pressures above $2$~GPa, computational studies support such an evolution of the ordering wave vector. In the following, we present total energy calculations based on density functional theory indicating that, indeed, there is a trend towards pressure-induced small-$Q$ AFM phases and that several small-$Q$ phases are nearly degenerate.

\begin{figure}[!htb]
\begin{center}
\includegraphics[width=8.1cm]{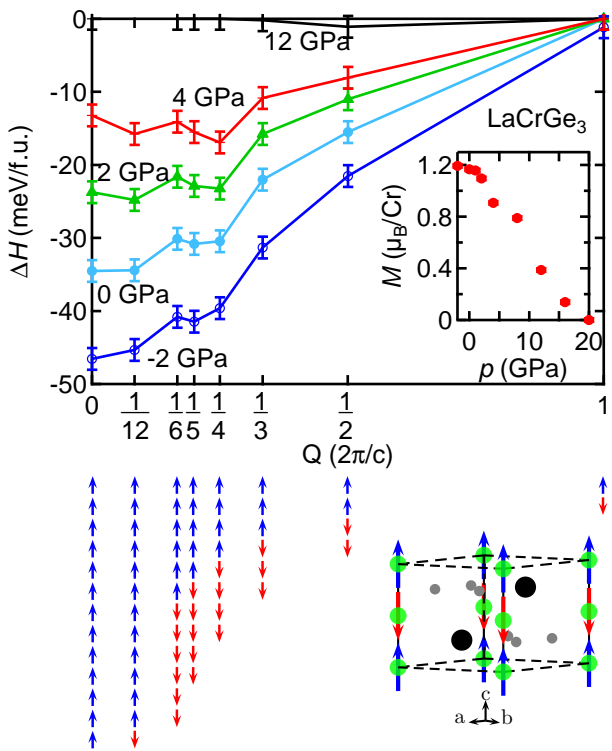}
\caption{\label{fig:LDA}Calculated enthalpy difference $\Delta H=H_{Q}-H_{\textrm{nonmagnetic}}$ as a function of the magnetic wave vector $Q$ for different pressures. The magnetic moment is not fixed in the calculation and the inset shows the pressure dependence of the calculated averaged moment for the most stable state. When a magnetic phase is not stable, the calculated moment is zero so that $H_{Q}=H_{\textrm{nonmagnetic}}$. The error bar is estimated from calculating the FM state enthalpy with various unit cell sizes~\cite{SupplMat}. The Cr moment of a simple antiferromagnetic state ($Q=2\pi/c$) is represented by arrows in the crystallographic unit cell, and the red and blue arrows for various $Q$~vectors are shown.}
\end{center}
\end{figure}

Figure~\ref{fig:LDA} shows the calculated enthalpy difference $\Delta H=H_{Q}-H_{\textrm{nonmagnetic}}$ as a function of the magnetic wave vector $Q$ along the $c$~axis for different pressures. Because of the rapid increase of computational resources and time with the system size (number of atoms), we limited our calculations to $Q=1/12$ reciprocal lattice units (r.l.u.), which contains 120 atoms in the unit cell. We can see clearly from Fig.~\ref{fig:LDA} that at $0$~GPa, the ferromagnetic ($Q=0$) state is stable. With increasing pressure, the FM state becomes less stable compared with states with small but finite $Q$ values. At $2$~GPa, the various states with $Q<1/4$~r.l.u. are nearly degenerate. First-principles density functional theory results are at $0$~K so that a small thermal smearing out of the enthalpy levels at finite temperature would make these states degenerate. These results are in agreement with a continuous evolution from $Q=0$ to $Q>0$ at finite temperature (near the Lifshitz point), whereas, at $0$~K, the wave vector may jump discontinuously to a finite value. At $2$~GPa, we found that $Q$ is at most $1/4$~r.l.u. whereas it could become $1/2$~r.l.u. at $12$~GPa, although a nonmagnetic ground state is within the error of the calculations (experimentally, we did not detect any anomaly above $5.3$~GPa). The calculations support very well the basic ideas that (i) the AFM$_{Q}$'s have long wavelength, (ii) the value of $Q\gtrsim0$ can change with pressure and temperature, and (iii) the Cr moment does not change significantly at the FM-AFM$_{Q}$ transition (inset of Fig.~\ref{fig:LDA}).

The observed phase diagram of LaCrGe$_3$, where the ferromagnetic ordering is driven to a QPT with a clean tuning parameter such as pressure, agrees with many of the recent predictions from the theories of quantum criticality. This phase diagram is consistent with the interesting prediction that modulated magnetic phases with small finite wave vectors should appear. This scenario is based on soft particle-hole excitations, which are always present in metals~\cite{Belitz2005PRL}. The soft modes couple to the magnetization and their correlation functions diverge in the limit of vanishing wave vector $Q$. One possible outcome is that the PM-FM transition becomes of the first order near $T=0$~K as observed in several compounds~\cite{Uemura2007Nature,Levy2007NP,Taufour2010PRL}. Another possibility, which seems revealed in LaCrGe$_3$, is that the ground state of the system is a modulated magnetic phase with a small-$Q$ wave vector that can vary with the tuning parameter such as spin-density wave and spiral phases~\cite{Belitz1997PRB,Chubukov2004PRL,Conduit2009PRL,Karahasanovic2012PRB,Thomson2013PRB,Pedder2013PRB,AbdulJabbar2015NP}.

Phases with small ordering vectors can also arise from competing exchange interactions between local moments~\cite{Bak1980PRB} as used to explain the complex magnetic structures of some rare-earth metals~\cite{Bak1980PRB,Elliott1961PR}. However, the applicability of such model to LaCrGe$_3$ is unclear because of the itinerant nature of the magnetism~\cite{Lin2013PRB,Bie2007CM}. In itinerant systems, spin density waves with a long period can form due to Fermi surface nesting at a small propagation vector near a ferromagnetic instability~\cite{Honda2006JPCM}. In this scenario, the nature of the QPT relies on detailed band-structure effects and seems less generic than the scenario based on soft particle-hole excitations~\cite{Belitz2005PRL}. In fact, determining whether the modulated magnetic phase is driven by quantum fluctuations or by electronic band dispersions is difficult~\cite{Conduit2009PRL}. So far, most materials studied with a clean tuning parameter such as pressure have a complex magnetism. In the helimagnet MnSi, partial long-range order was observed under pressure~\cite{Pfleiderer2004Nature}. In another helimagnet MnP, another magnetic phase also appears under pressure with superconductivity near the antiferromagnetic quantum critical point~\cite{Cheng2015PRL}. Kondo systems such as CeRuPO~\cite{Kotegawa2013JPSJ,Lengyel2015PRB} or the induced moment magnet PrPtAl~\cite{AbdulJabbar2015NP} have shown evidence for a modulated phase. For those systems, a detailed comparison with band-structure calculations will be difficult since magnetism originates from rare-earth elements. LaCrGe$_3$, a simple $3d$ electrons system, with a simple ferromagnetic structure at ambient pressure, provides a unique opportunity for a quantitative comparison with band-structure calculations. Our results indicate that band-structure effects provide a trend toward the formation of small-$Q$ AFM phases in LaCrGe$_3$. The near degeneracy of different $Q$~phases provides a relatively flat energy landscape allowing quantum fluctuations to play an important role in the selection of modulated phases.

To conclude, the discovery of a new magnetic phase in place of a ferromagnetic quantum critical point in LaCrGe$_3$ provides a clear example of avoided quantum criticality when a ferromagnetic transition is suppressed by a clean tuning parameter such as pressure. Our transport, thermodynamic and microscopic measurements under pressure strongly establish this phase diagram which is compatible with band-structure calculations as well as general predictions based on soft excitations. The results presented here pose a formidable challenge for both models as to provide enough quantitative predictions allowing us to distinguish band-structure or quantum fluctuation effects. Experimentally, the exact nature of the $Q$~vector in the new magnetic phase at high pressure as well as its temperature and pressure evolution remain to be determined. Theoretically, it needs to be clarified when the ferromagnetic-paramagnetic transition becomes of the first order and when a new modulated phase will appear.

\begin{acknowledgments}
We would like to thank V.~G.~Kogan, A.~Kreyssig, P.~Kumar, and D.~K.~Finnemore for useful discussions, as well as D.~Belitz, T.~R.~Kirkpatrick, F.~Kr\"uger, and A.~G.~Green for their critical reading of the manuscript.
This work was supported by the Materials Sciences Division of the Office of Basic Energy Sciences of the U.S. Department of Energy. Part of this work was performed at the Ames Laboratory, US DOE, under Contract No. DE-AC02-07CH11358. Magnetization measurements under pressure (V.~T.) were supported by Ames Laboratory's laboratory-directed research and development (LDRD) funding. X.~L., E.~D.~M., and H.~K. were supported by the AFOSR-MURI Grant No. FA9550-09-1-0603.
\end{acknowledgments}

\end{document}